\documentclass[twocolumn,superscriptaddress,aps,pra]{revtex4}

\usepackage{graphicx}

\usepackage{color}

\begin{document}
\title{Effect of laser frequency fluctuation on the decay rate of Rydberg coherence}
\author{Bongjune Kim}
\affiliation{Department of Physics, National Tsing Hua University, Hsinchu 30013, Taiwan}
\author{Ko-Tang Chen}
\affiliation{Department of Physics, National Tsing Hua University, Hsinchu 30013, Taiwan}
\author{Chia-Yu Hsu}
\affiliation{Department of Physics, National Tsing Hua University, Hsinchu 30013, Taiwan}
\author{Shih-Si Hsiao}
\affiliation{Department of Physics, National Tsing Hua University, Hsinchu 30013, Taiwan}
\author{Yu-Chih Tseng}
\affiliation{Department of Physics, National Tsing Hua University, Hsinchu 30013, Taiwan}
\author{Chin-Yuan Lee}
\affiliation{Department of Physics, National Tsing Hua University, Hsinchu 30013, Taiwan}
\author{Shih-Lun Liang}
\affiliation{Department of Physics, National Tsing Hua University, Hsinchu 30013, Taiwan}
\author{Yi-Hua Lai}
\affiliation{Department of Physics, National Tsing Hua University, Hsinchu 30013, Taiwan}
\author{Julius Ruseckas}
\affiliation{Institute of Theoretical Physics and Astronomy, Vilnius University, Saul\.{e}tekio 3, LT-10222 Vilnius, Lithuania}
\author{Gediminas Juzeli\= unas}
\affiliation{Institute of Theoretical Physics and Astronomy, Vilnius University, Saul\.{e}tekio 3, LT-10222 Vilnius, Lithuania}
\author{Ite A. Yu}
\email{yu@phys.nthu.edu.tw}
\affiliation{Department of Physics, National Tsing Hua University, Hsinchu 30013, Taiwan}
\affiliation{Center for Quantum Technology, Hsinchu 30013, Taiwan}

\date{\today}
\begin{abstract}
The effect of electromagnetically induced transparency (EIT) combined with Rydberg-state atoms provides high optical nonlinearity to efficiently mediate the photon-photon interaction. However, the decay rate of Rydberg coherence, i.e., the decoherence rate, plays an important role in optical nonlinear efficiency, and can be largely influenced by laser frequency fluctuation. In this work, we carried out a systematic study of the effect of laser frequency fluctuation on the decoherence rate. We derived an analytical formula that quantitatively describes the relationship between the decoherence rate and laser frequency fluctuation. The formula was experimentally verified by using the $\Lambda$-type EIT system of laser-cooled $^{87}$Rb atoms, in which one can either completely eliminate or controllably introduce the effect of laser frequency fluctuation. We also included the effect of Doppler shift caused by the atomic thermal motion in the formula, which can be negligible in the $\Lambda$-type EIT experiment but significant in the Rydberg-EIT experiment. Utilizing the atoms of 350~$\mu$K, we studied the decoherence rate in the Rydberg-EIT system involving with the state of $|32D_{5/2}\rangle$. The experimental data are consistent with the predictions from the formula. We were able to achieve a rather low decoherence rate of $2\pi\times$48 kHz at a moderate coupling Rabi frequency of $2\pi\times$4.3 MHz.
\end{abstract}

\maketitle
\newcommand{\FigOne}{
	\begin{figure}[t]
	\centering\includegraphics[width=0.43\textwidth]{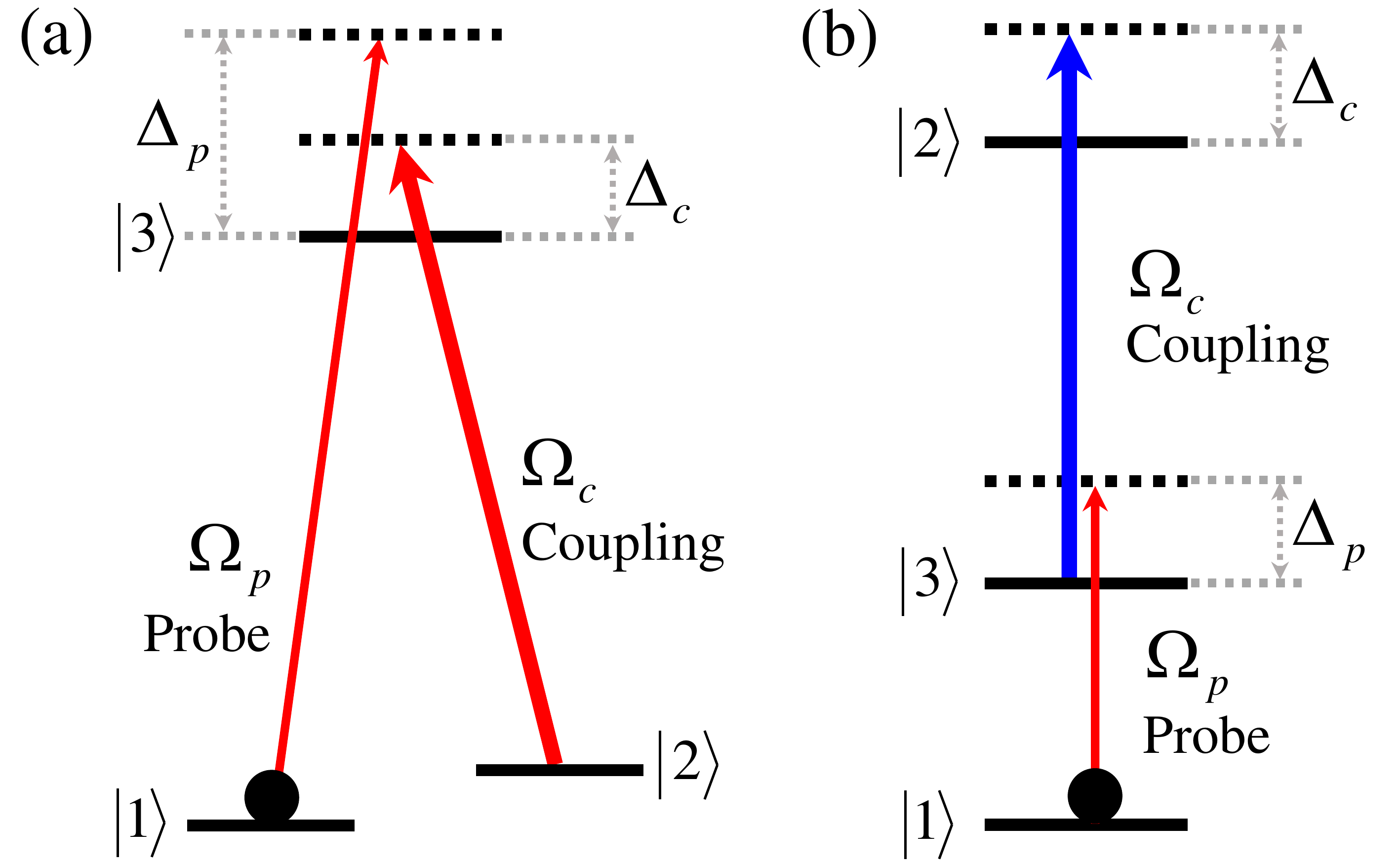}
	\caption{Relevant energy levels and laser excitations in the $\Lambda$-EIT system are shown in (a) and those in the Rydberg-EIT system are shown in (b). We employed laser-cooled $^{87}$Rb atoms in the experiment. In (a), states $|1\rangle$ and $|2\rangle$ correspond to the ground states $|5S_{1/2},F=1,m_F=1\rangle$ and $|5S_{1/2},F=2,m_F=1\rangle$, and state $|3\rangle$ corresponds to the excited state $|5P_{3/2},F=2, m_F=2\rangle$. In (b), states $|1\rangle$, $|2\rangle$, and $|3\rangle$ correspond to the ground state $|5S_{1/2},F=2,m_F=2\rangle$, the Rydberg state $|32D_{5/2},m_J=5/2\rangle$, and the excited state $|5P_{3/2},F=3,m_F=3\rangle$, respectively. 
	}
	\label{fig:Transitions}
	\end{figure}
}
\newcommand{\FigTwo}{
	\begin{figure}[t]
	\centering\includegraphics[width=0.48\textwidth]{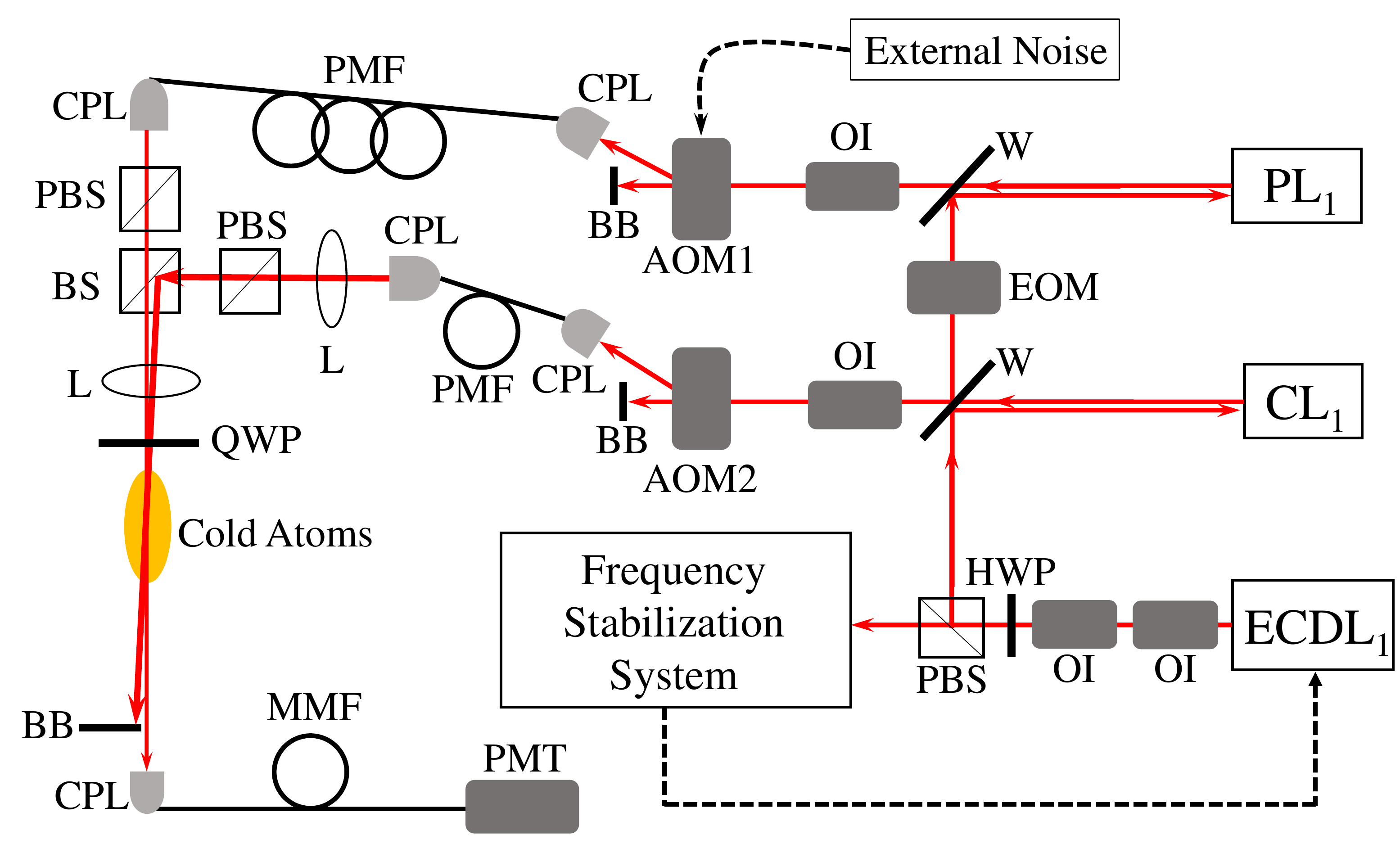}
	\caption{Experimental setup of the $\Lambda$-EIT study. ECDL: external-cavity diode laser, PL: probe laser, CL: coupling laser, EOM: electro-optic modulator, OI: optical isolator, HWP: half-wave plate; AOM: acousto-optic modulator, BB: beam block; CPL: optical fiber coupler, PMF: polarization-maintained optical fiber, PBS: polarizing beam splitter, BS: beam splitter ($T/R = 10/90$), W: window, L: lens, QWP: quarter-wave plate, MMF: multi-mode optical fiber, and PMT: photo-multiplier tube. 
	}
	\label{fig:SetupLEIT}
	\end{figure}
}
\newcommand{\FigThree}{
	\begin{figure}[t]
	\centering\includegraphics[width=0.35\textwidth]{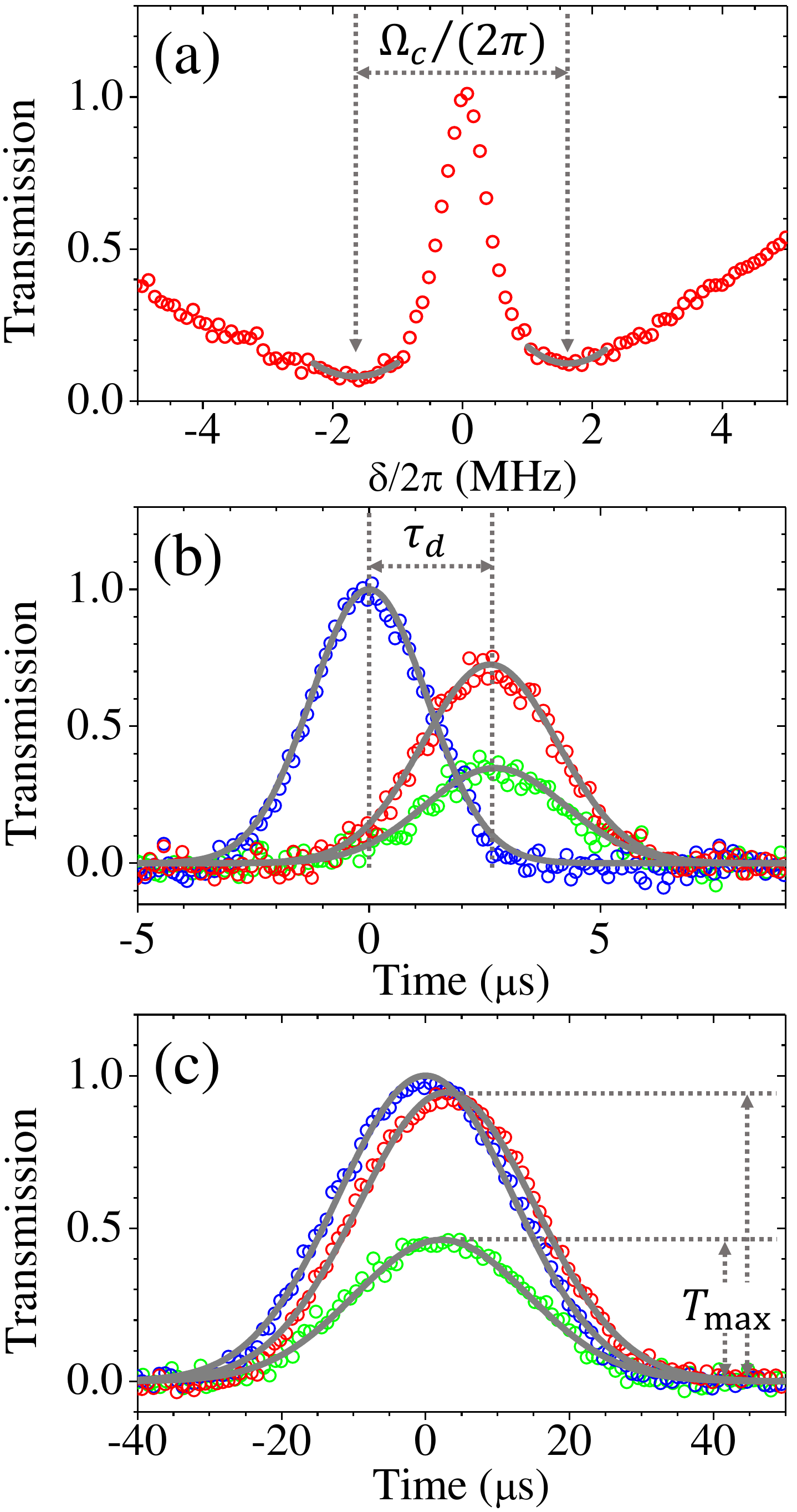}
	\caption{Determination of experimental parameters in the $\Lambda$-EIT study. (a) EIT spectrum represented by red circles was measured with an intentionally-reduced optical depth. The separation distance between two transmission minima determines the coupling Rabi frequency $\Omega_c$ = 0.54$\Gamma$. (b) Slow light data of short probe pulses at $\Omega_c$ = 0.54$\Gamma$. The input pulse is represented by blue circles. The output pulses under the frequency fluctuation $\Gamma_f/(2\pi)$ = 0 and 280~kHz are represented by red and green circles, respectively. Gray lines are the Gaussian best fits to identify peak positions of the pulses. The delay time between the input and output pulses determines the optical depth $\alpha$ = 29. (c) Slow light data of long probe pulses at $\Omega_c$ = 0.54$\Gamma$ and $\alpha$ = 29. Legends are the same as those in (b). The Gaussian best fits identify peak transmissions of the output pulses, which determine the decoherence rates $\gamma = 2.9\times10^{-4}\Gamma$ (red) and $3.9\times10^{-3}\Gamma$ (green).
	}
	\end{figure}
}
\newcommand{\FigFour}{
	\begin{figure}[t]
	\centering\includegraphics[width=0.38\textwidth]{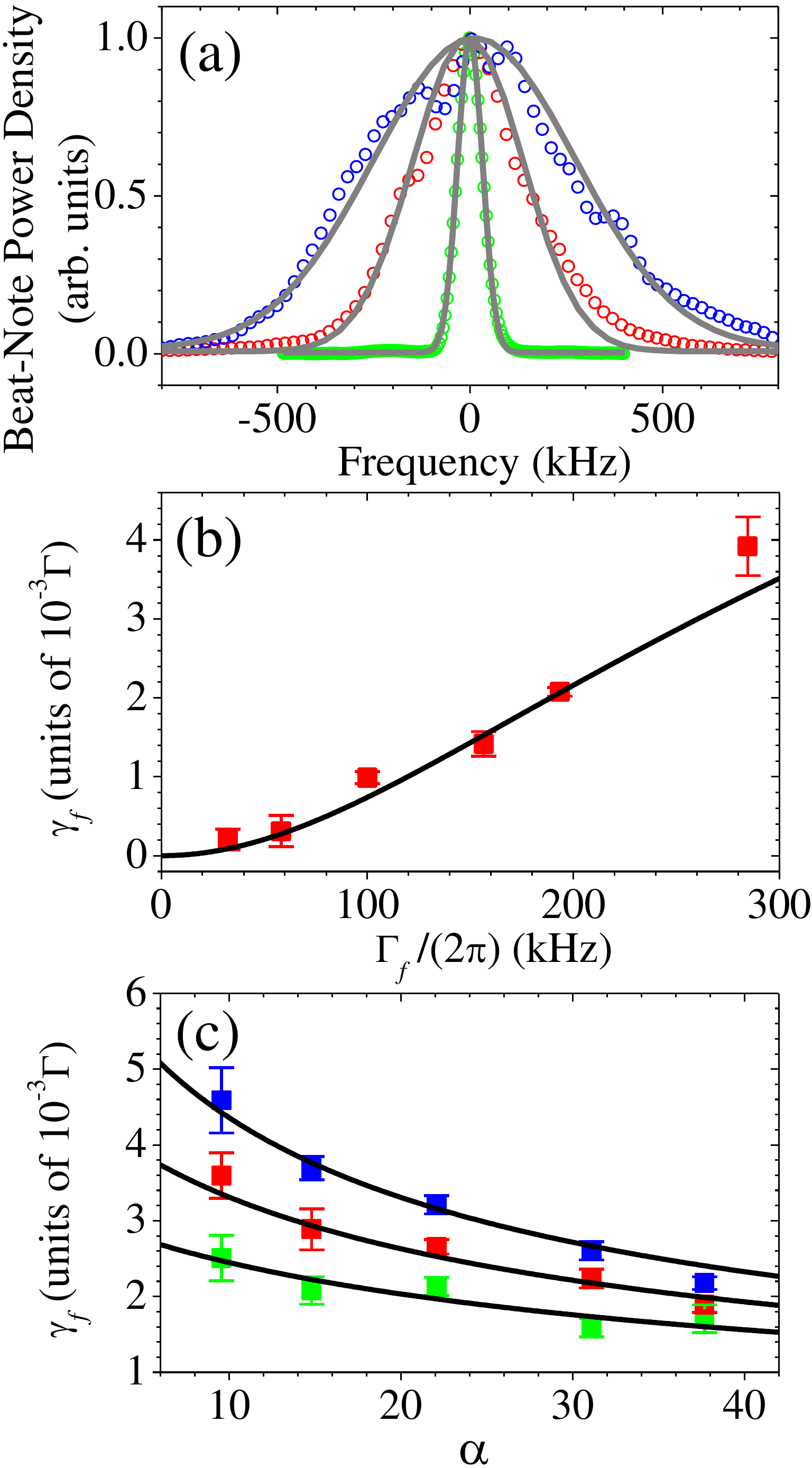}
	\caption{(a) Representative spectra of beat notes between the first-order and zeroth-order beams of the AOM1 in Fig.~2 under different noise amplitudes, $\Gamma_f$. Blue, red, and green circles are the spectra, and gray lines are the Gaussian best fits. (b) The frequency fluctuation-induced decoherence rate $\gamma_f$ as a function of $\Gamma_f$ at $\Omega_c$ = 0.54$\Gamma$ and $\alpha$ (optical depth) = 29. Red squares are the experimental data. (c) $\gamma_f$ as a function of $\alpha$ at $\Omega_c$ = 0.44$\Gamma$. Blue, red, and green squares are the experimental data measured with $\Gamma_f/(2\pi)$ = 220, 180, and 150 kHz, respectively. In (b) and (c), blacks lines are the theoretical predictions according to Eq.~(\ref{eq:gammaf}).
	}
	\end{figure}
}
\newcommand{\FigFive}{
	\begin{figure}[t]
	\centering\includegraphics[width=0.48\textwidth]{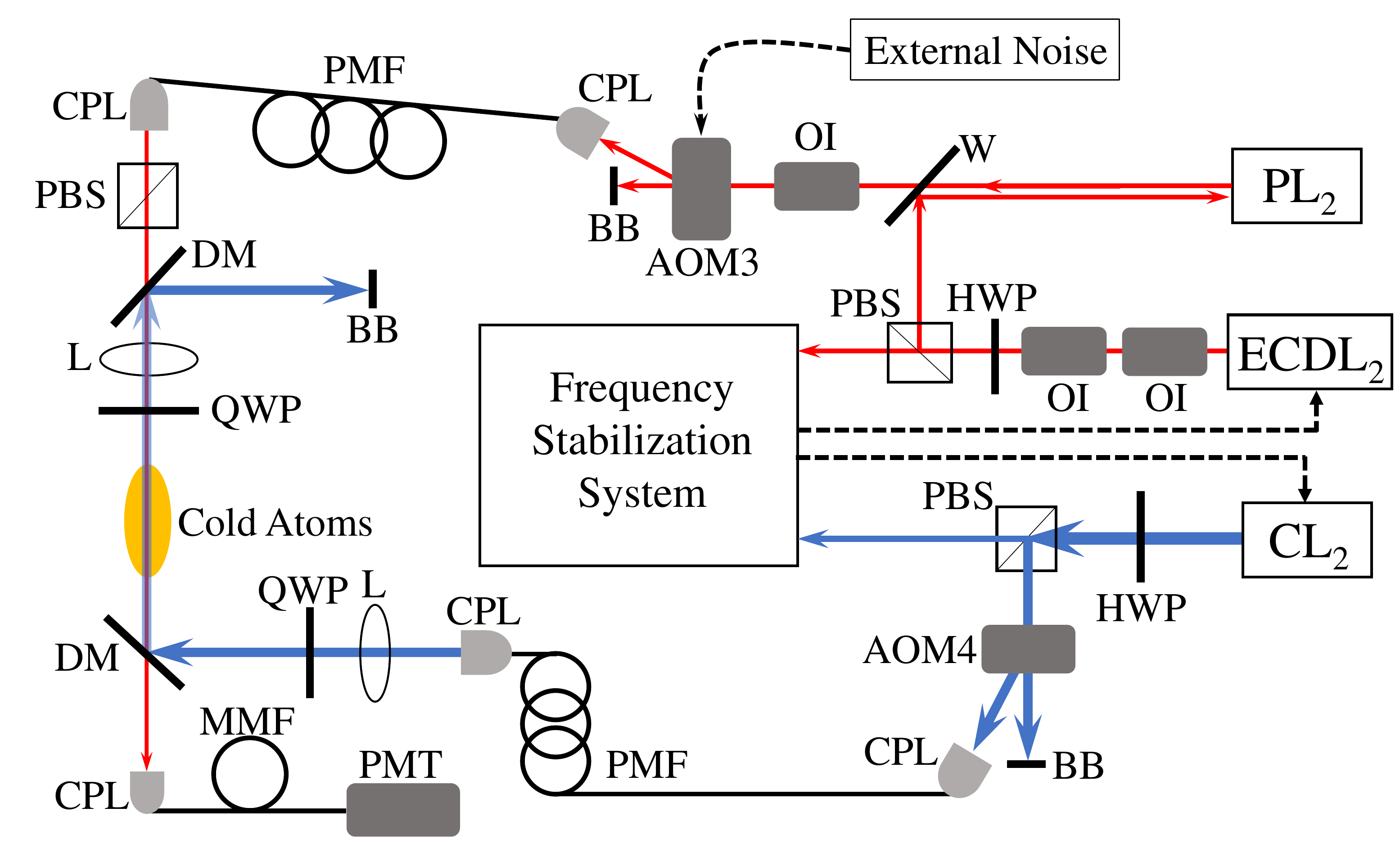}
	\caption{Experimental setup of the Rydberg-EIT study. ECDL: external-cavity diode laser  (Toptica DLC DL pro), PL: probe laser, CL: coupling laser (Toptica TA-SHG pro), OI: optical isolator, W: window, HWP: half-wave plate, PBS: polarizing beam splitter, AOM: acousto-optic modulator, BB: beam block, CPL: optical fiber coupler, PMF: polarization-maintained optical fiber, DM: dichroic mirror, L: lens, QWP: quarter-wave plate, MMF: multi-mode optical fiber, and PMT: photo-multiplier tube.
	}
	\end{figure}
}
\newcommand{\FigSix}{
	\begin{figure}[t]
	\centering\includegraphics[width=0.38\textwidth]{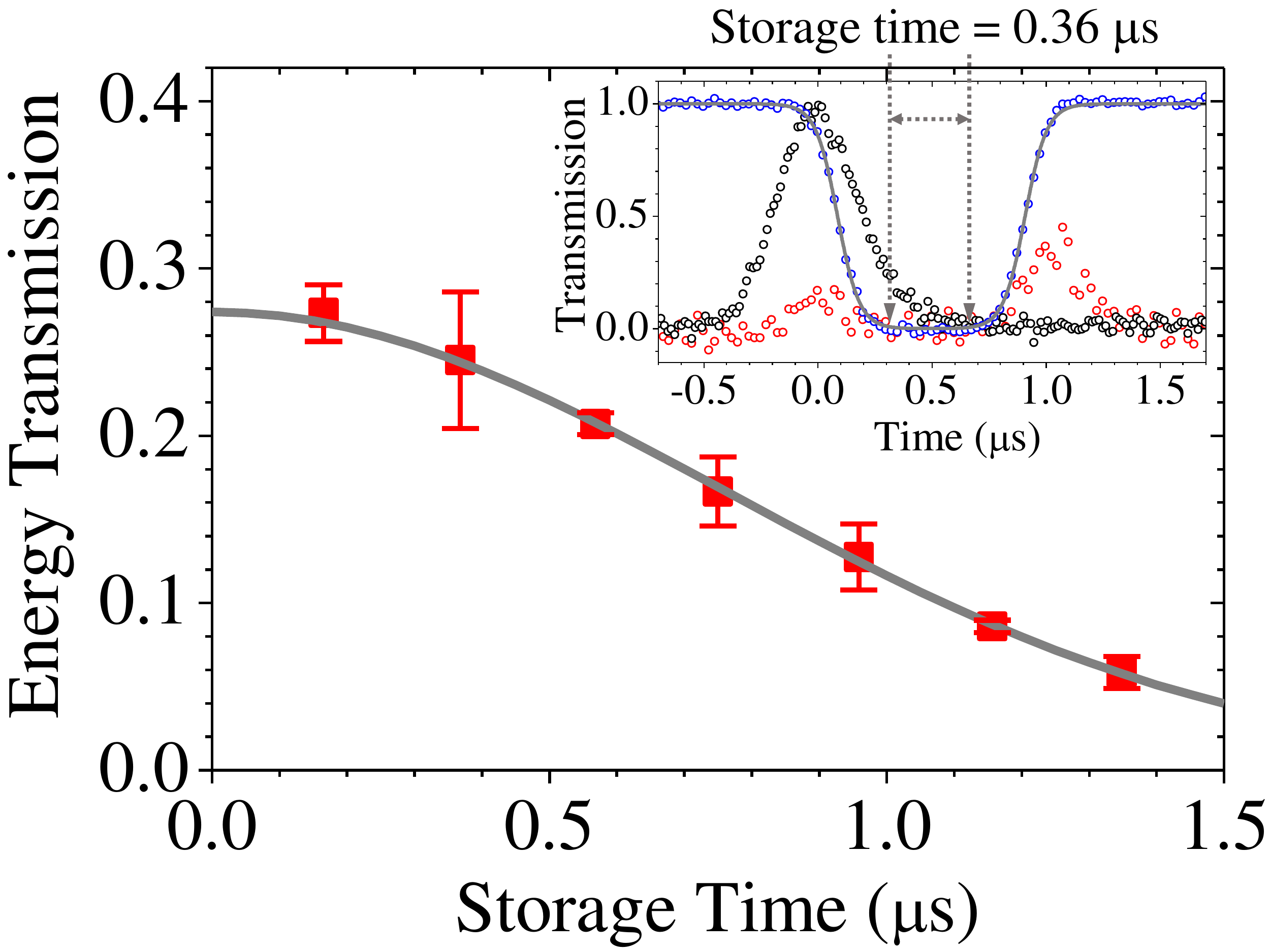}
	\caption{The main plot shows retrieval efficiency (ratio of output to input energies) of the probe pulse as a function of storage time in the Rydberg-EIT system. Red squares are the experimental data of retrieval efficiency measured with $\Omega_c =$ 1.6$\Gamma$ and $\alpha=30$. Gray line is the Gaussian best fit. The coherence time or $e^{-1}$ decay time of the best fit is 1.1~$\mu$s, corresponding to the atom temperature of 350~$\mu$K in the experiment. The inset shows representative data of storage and retrieval. Black and red circles represent the signals of the input and output probe pulses. Blue circles represent the signal of the coupling field. Gray line is the best fit of the sum of two hyperbolic-tangent functions, describing falling and rising behaviors of the coupling field. We define the storage time as the interval that the coupling field is completely off.
	}
	\end{figure}
}
\newcommand{\FigSeven}{
	\begin{figure}[t]
	\centering\includegraphics[width=0.35\textwidth]{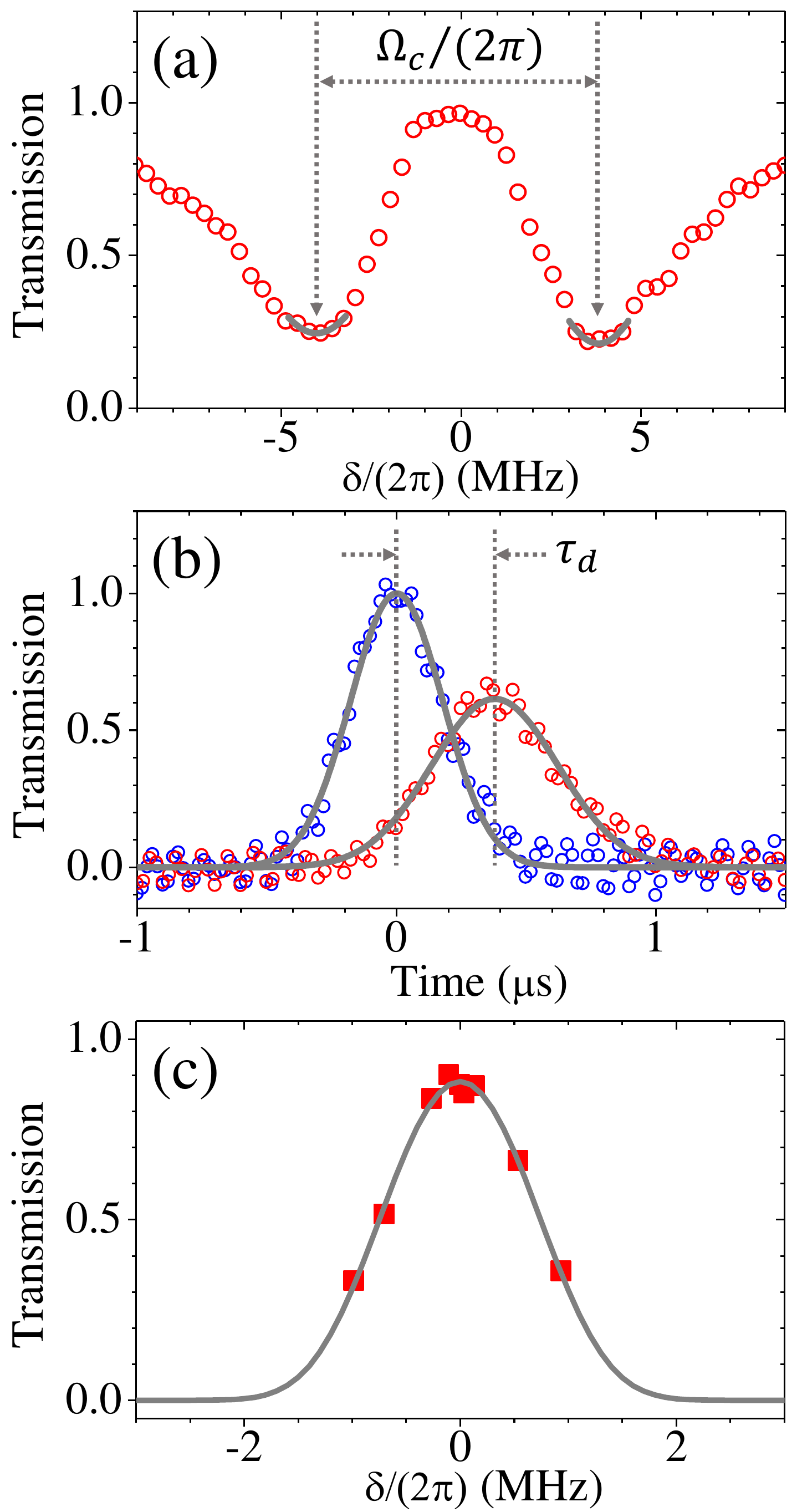}
	\caption{Determination of experimental parameters in the Rydberg-EIT study. (a) EIT spectrum represented by red circles was measured with an intentionally-reduced optical depth. The separation distance between two transmission minima determines the coupling Rabi frequency $\Omega_c$ = 1.3$\Gamma$. (b) Slow light data of a short probe pulse at $\Omega_c$ = 1.3$\Gamma$. Blue and red circles are the input and output pulses, respectively. Gray lines are the Gaussian best fits to identify peak positions of the pulses. The delay time between the input and output pulses determines the optical depth $\alpha$ = 25. (c) Transmission of the output probe pulse as a function of the two-photon detuning at $\Omega_c$ = 1.3$\Gamma$ and $\alpha$ = 25. The $e^{-1}$ full width of the input pulse was 7.0~$\mu$s. Red squares are the experimental data. Gray line is the best fit  which determines the decoherence rate $\gamma = 4.6$$\times$$10^{-3}$$\Gamma$.
	}
	\end{figure}
}
\newcommand{\FigEight}{
	\begin{figure}[t]
	\centering\includegraphics[width=0.4\textwidth]{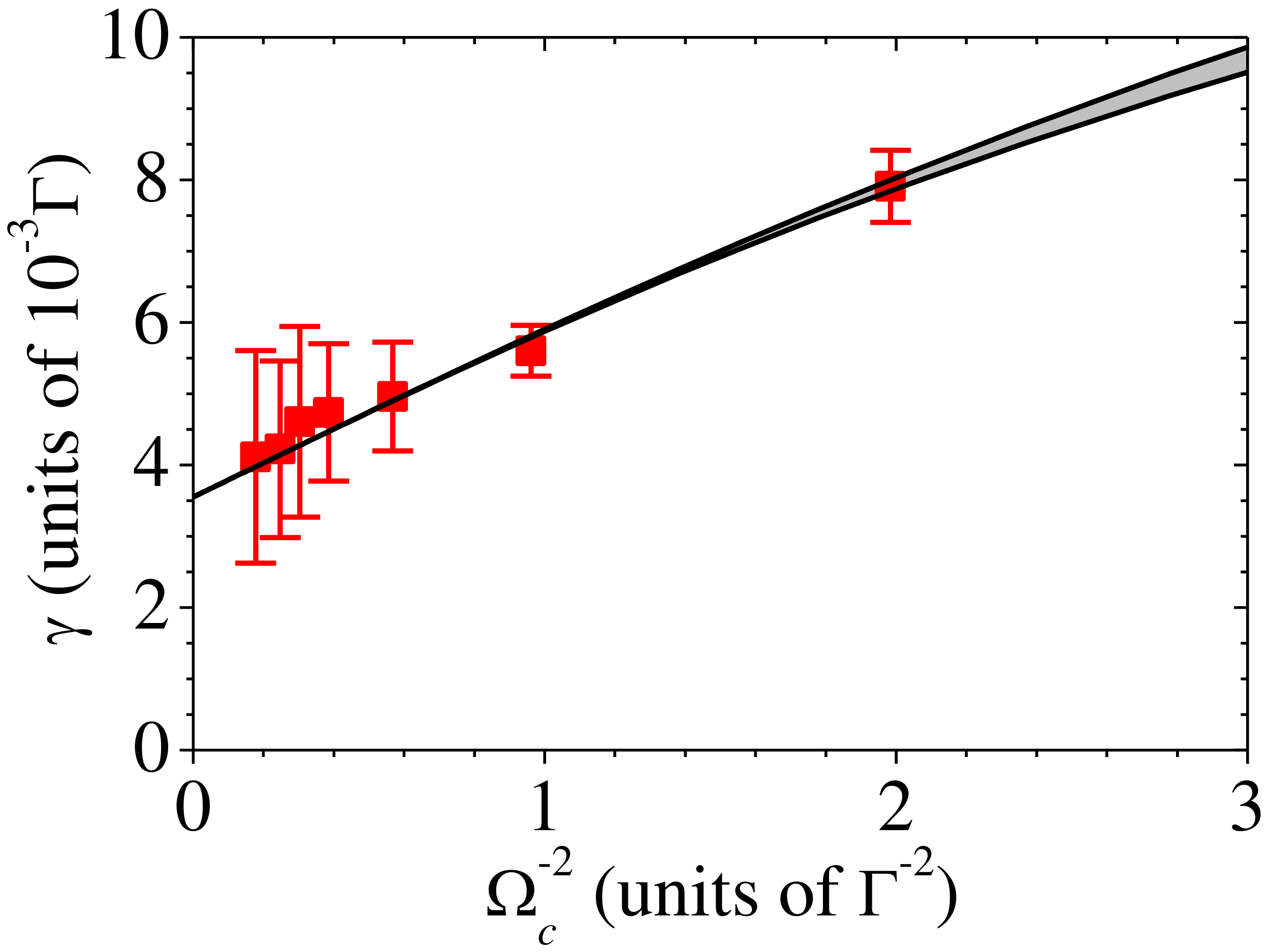}
	\caption{Decoherence rate $\gamma$ in the Rydberg-EIT system as a function of $\Omega_c^{-2}$. Red squares are the averages of data taken at different optical depths ($\alpha$) of 15$\sim$26. Two black lines, corresponding to $\alpha$ = 15 (upper) and 26 (lower), are the theoretical predictions of Eq.~(\ref{eq:gamma}). The predictions were calculated with $\gamma_0 =$ 3.6$\times$$10^{-3}$$\Gamma$ or $2\pi$$\times$22 kHz, $\Gamma_D$ =  $2\pi$$\times$200 kHz, and $\Gamma_f$ =  $2\pi$$\times$210 kHz (see text for how to determine these values).
	}
	\end{figure}
}
\newcommand{\FigNine}{
	\begin{figure}[t]
	\centering\includegraphics[width=0.4\textwidth]{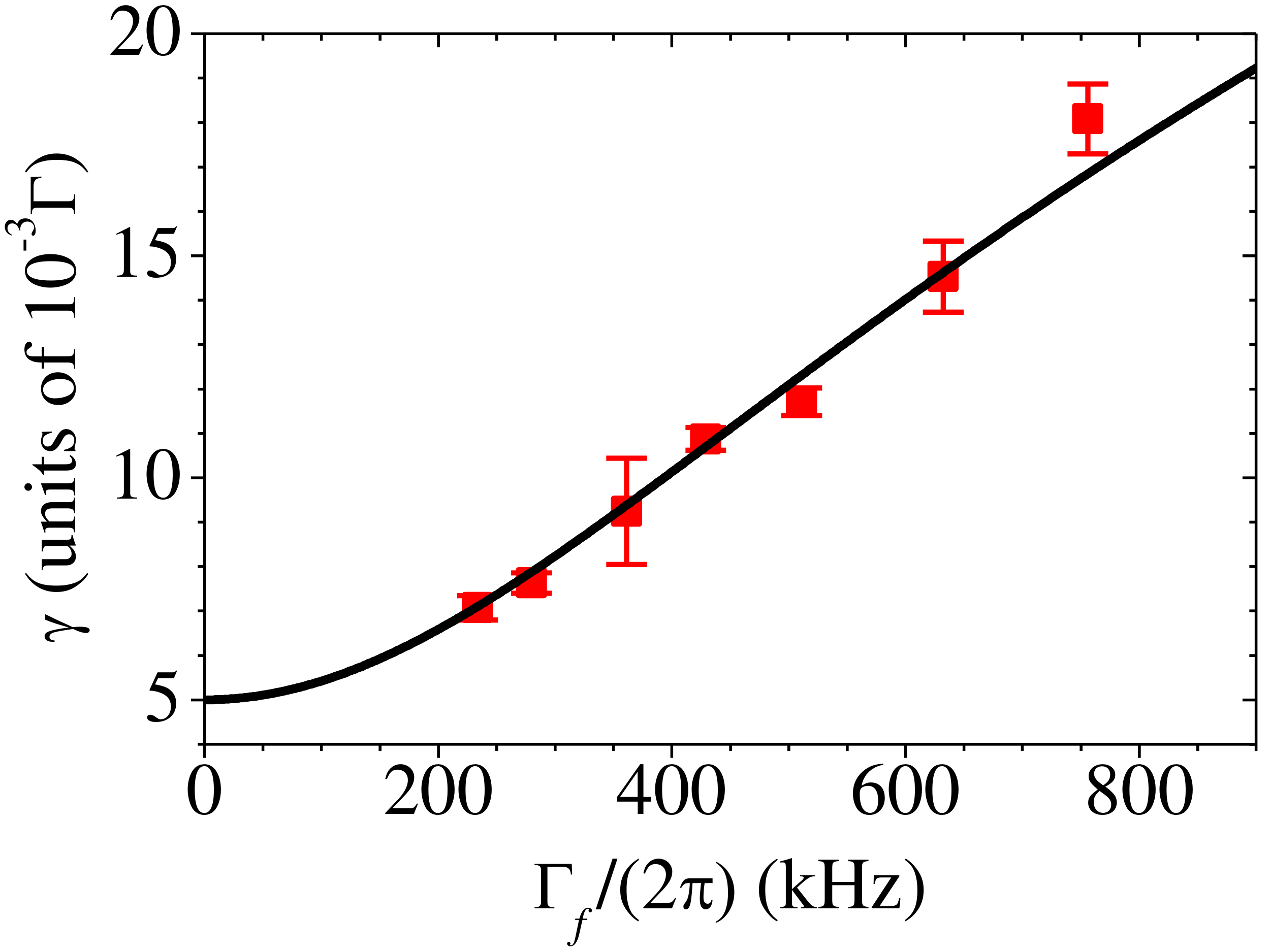}
	\caption{Decoherence rate $\gamma$ in the Rydberg-EIT system as a function of frequency fluctuation $\Gamma_f$. In the measurement, $\Omega_c$ = 0.83$\Gamma$ and $\alpha$ (optical depth) = 18. Red squares are the experimental data, and black line is the theoretical prediction according to Eq.~(\ref{eq:gamma}).
	}
	\end{figure}
}

\section{Introduction}

Utilizing the strong dipole-dipole interaction (DDI) between Rydberg-state atoms in the applications of quantum information processing, such as realization of quantum logic gates~\cite{Saffman2010, Browaeys2010, Lukin2018}, generation of single photons~\cite{Saffman2002, Kuzmich2016}, and quantum simulations~\cite{Zoller2010, Ahn2018}, is of great interest currently. These applications are made possible by the DDI-induced blockade effect, the phenomenon that multiple excitations to a Rydberg state within the blockade radius is strongly suppressed~\cite{Pfau2007, Comparat2010, Fleischhauer2011, Kuzmich2012, Hofferberth2016, A1, A2, A3, A4, A5}. On the other hand, the effect of electromagnetically induced transparency (EIT) provides high optical nonlinearity~\cite{Fleischhauer2005, OurPRA2014, OurPRL2016}. Hence, the EIT effect combined with Rydberg-state atoms can efficiently mediate the photon-photon interaction, offering a powerful tool for quantum information manipulation with photons~\cite{Adams2010, Lukin2012, B1, B2, B3, B4, B5, B6}. Furthermore, the storage of light based on the EIT effect can prolong the atom-photon or photon-photon interaction time~\cite{OurPRL2006, OurPRL2009, Durr2014, Durr2016, Vuletic2016, Storage2016, Ruseckas2017, Pohl2018}. Assisted by long lifetimes of Rydberg states, all-optical switching or cross-phase shift with single photons and single-photon subtraction have been demonstrated with the light-storage scheme in the Rydberg-EIT system~\cite{Durr2014, Durr2016, Pohl2018}.

Similar to the Raman or ground-state coherence being the coherence between the two ground states in the $\Lambda$-type EIT (abbreviated as $\Lambda$-EIT) system depicted in Fig.~1(a)~\cite{Fleischhauer2000, Juzeliunas2002}, the Rydberg coherence is the coherence between the ground and Rydberg states in the Rydberg-EIT system depicted in Fig.~1(b). The decay rate of Rydberg coherence, i.e., the decoherence rate, can greatly influence the optical nonlinear efficiency of the Rydberg-EIT effect~\cite{Adams2008, Leseleuc2018, Pritchard2018, stability2018, C1}. Because laser frequency fluctuation increases the decoherence rate, it also deteriorates the nonlinear efficiency. Hence, laser frequency fluctuation can be a problem in the high-fidelity low-loss quantum processes utilizing the Rydberg-EIT scheme.

The $\Lambda$-EIT system in Fig.~1(a) is driven by the probe and coupling fields in the $\Lambda$-type configuration. The frequency difference between the probe and coupling lasers determines the two-photon detuning. One can employ the phase-lock or injection-lock scheme to completely eliminate the laser frequency fluctuation from this frequency difference. Thus, the EIT resonant condition is stabilized to a high degree \cite{OurPRL2013}, and the laser frequency fluctuation is not a problem in the $\Lambda$-EIT experiment. On the other hand, the Rydberg-EIT system in Fig.~1(b) is driven by the probe and coupling fields in the ladder-type configuration. The sum of the probe and coupling frequencies determines the two-photon detuning. The schemes of reference cavities, high-resolution wavemeters, EIT spectroscopy, etc.\ were employed for the stabilization of laser frequencies in the Rydberg-EIT experiments~\cite{cavity2017, wavemeter2011, Adams2009}. However, the laser frequency fluctuation resulting from any stabilization method still contributes to this frequency sum, and makes the experimental condition deviate from the EIT resonance. Thus, the laser frequency fluctuation is an unavoidable problem in the Rydberg-EIT experiment.

In Refs.~\cite{Xiao1} and \cite{Xiao2}, the authors developed a theory for Doppler-broadened EIT media and also considered the effect of laser linewidth. They experimentally studied the theory in the ladder-type system with a room-temperature atomic vapor. In Ref.~\cite{Xiao3}, L\"{u} \emph{et al}. experimentally investigated how the peak transmission of the $\Lambda$-EIT system is influenced by frequency fluctuation of the coupling field with a room-temperature atomic vapor. However, the effect of the frequency fluctuation was phenomenologically introduced in this reference. In Refs.~\cite{Adams2008, Adams2009}, the authors used the similar way to introduce the laser frequency fluctuation to the decoherence rate in their Rydberg-EIT experiments. Here, we carried out a systematic study of the effect of laser frequency fluctuation on the decoherence rate. We derived a formula that quantitatively describes the relationship between the decoherence rate and laser frequency fluctuation, and explicitly shows the roles of the coupling Rabi frequency and the optical depth in the relationship. The derived formula was experimentally verified in both of the $\Lambda$-EIT and Rydberg-EIT systems with cold $^{87}$Rb atoms.

\FigOne

This article is organized as the followings. In Sec.~II, we will derive an analytical formula showing the decoherence rate as a function of the laser frequency fluctuation. Since the Doppler shift caused by the atomic thermal motion is not negligible in our Rydberg-EIT system, we also include the effect of the Doppler shift in the formula. In Sec.~III, we will report the test result on the validity of the formula in the $\Lambda$-EIT system. Figure~3 illustrates the methods that determine the coupling Rabi frequency, optical depth, and decoherence rate. Figure~4 demonstrates that experimental data of the decoherence rate in the $\Lambda$-EIT system are consistent with predictions from the formula. In Sec.~IV, we will report the study on the decoherence rate in the Rydberg-EIT system. The Rydberg state $|32D_{5/2}\rangle$ was selected in the study to avoid the DDI effect~\cite{Li2016}. Figure~6 provides the information about the atom temperature. The purpose of Fig.~7 is the same as that of Fig.~3. Figures~8 and 9 demonstrate that experimental data of the decoherence rate in the Rydberg-EIT system are consistent with predictions from the formula. Finally, we will make a conclusion in Sec.~V.


\section{Theoretical Model}

Considering the two transition diagrams in Fig.~1, we derive an analytic formula that relates the decoherence rate of $\rho_{21}$ to the laser frequency fluctuation.  In Fig.~1(a), $\rho_{21}$ is the coherence between the two ground states. In Fig.~1(b), $\rho_{21}$ is the coherence between the ground state and the Rydberg state. The logic behind the derivation is the following. Although the laser frequencies are locked to the resonance frequency of the two-photon transition, their fluctuations randomly induce two-photon detunings to the EIT system. The two-photon detuning leads to attenuation or loss of the probe field. The average value of attenuations of various two-photon detunings caused by the laser frequency fluctuation can be seen as the result of an effective decoherence rate. A larger amplitude of the laser frequency fluctuation represents a greater root-mean-square value of the two-photon detuning, which makes more attenuation, similar to a larger decoherence rate. Therefore, the decoherence rate can be expressed as a function of the fluctuation amplitude.

We employed the optical Bloch equation (OBE) for the density-matrix operator of the atomic ensemble and the Maxwell-Schr\"{o}dinger equation (MSE) for the probe field in the derivation, giving \cite{OurPRA2017}
\begin{eqnarray}
	\frac{\partial}{\partial t}\rho_{21} = 
		\frac{i}{2}\Omega_{c}\rho_{31} +i\delta\rho_{21} -\gamma_0\rho_{21}, \\
	\frac{\partial}{\partial t}\rho_{31} =
		\frac{i}{2}\Omega_{p} +\frac{i}{2}\Omega_{c}\rho_{21}
		+i\Delta_{p}\rho_{31} -\frac{\Gamma}{2}\rho_{31}, \\
	\frac{1}{c}\frac{\partial}{\partial t}\Omega_p
		+\frac{\partial}{\partial z}\Omega _p = i\frac{\alpha\Gamma}{2L}\rho_{31},
\end{eqnarray}
where $\rho_{21}$ and $\rho_{31}$ are the density-matrix elements, $\Omega_p$ and $\Omega_c$ denote probe and coupling Rabi frequencies, $\gamma_0$ is the decoherence rate, $\Gamma$ represents the spontaneous decay rate of the excited state $|3\rangle$ which is 2$\pi$$\times$ 6.1 MHz in our case, $\delta$ is the two-photon detuning of the Raman or Rydberg transition $|1\rangle$$\rightarrow$$|2\rangle$, $\Delta_p$ denotes the one-photon detuning of the probe transition  $|1\rangle$$\rightarrow$$|3\rangle$, and $\alpha$ and $L$ represent the optical depth (OD) and length of the medium.

To achieve the above OBE and MSE, we consider the weak probe field as a perturbation \cite{OurPRA2017}, and neglect the effect of dipole-dipole interaction among the Rydberg atoms. Only the slowly-varying amplitudes of the density-matrix elements and those of the probe and coupling Rabi frequencies remain in the equations. All parts of the equations, except $\delta$, are the same for both the $\Lambda$-EIT system shown in Fig.~1(a) and the Rydberg-EIT system shown in Fig.~1(b). The two-photon detuning is $\delta$ = $\Delta_p -\Delta_c$ for the situation shown in Fig.~1(a) and $\delta$ = $\Delta_p + \Delta_c$ for that in Fig.~1(b), where $\Delta_c$ is the one-photon detuning of the coupling transition.

To find the EIT spectral profile, we use Eqs.~(1) and (2) and obtain the following steady-state solution for $\rho_{31}$:
\begin{equation}
	\rho_{31} = 
		\frac{\delta + i\gamma_0}
		{\Omega_{c}^{2}/2-2(\Delta_p+ i\Gamma/2)(\delta + i\gamma_0)}\Omega_p.
\label{eq:rho31}
\end{equation}
The imaginary and real parts of $\rho_{31}$ determine the output transmission and phase shift of the probe field, respectively. We are only interested in the transmission. Under the typical EIT condition of $\Omega_c^2 \gg 2\gamma_0\Gamma$ and $\Omega_c^2 \gg 4 \delta\Delta_p$, the absorption cross section, $\sigma$, relates to the imaginary part of $\rho_{31}$ as the following:
\begin{equation}
	\sigma(\delta) = {\rm Im}\left[ \frac{\rho_{31}\Gamma}{\Omega_p} \right] 
		\approx \frac{2\gamma_0\Gamma}{\Omega_c^2} 
		+\frac{4 \Gamma^2 \delta^2}{\Omega_c^4}.
\label{eq:sigma}
\end{equation}
To obtain the steady-state output transmission of the probe field, we drop the time derivative term in Eq.~(3), and use the expression of $-[(2\gamma_0/\Omega_c^2) + (4\Gamma\delta^2/\Omega_c^4)]\Omega_p$ for $i\rho_{31}$ on the right-hand side of Eq.~(3). After Eq.~(3) is solved analytically, one arrives at the following output-to-input ratio or transmission of the probe field as a function of the two-photon detuning:
\begin{equation}
	 t(\delta) =  \frac{|\Omega_p(L)|^2}{|\Omega_p(0)|^2} = e^{-\alpha\sigma(\delta)}.
\label{eq:t}
\end{equation}

Nonzero two-photon detunings can exist in the frame of moving atoms due to Doppler shift. A higher velocity results in a larger two-photon detuning, $\delta_D$. Since the distribution of the atom velocity is a Gaussian function, the average of the absorption cross section due to the atomic motion is given by \cite{OurPRA2011}
\begin{eqnarray}
	\bar{\sigma}(\delta) &=&
		\int_{-\infty}^{\infty} d\delta_D 
		\frac{e^{-(\delta_D/\Gamma_D)^2} }{\sqrt{\pi}\Gamma_D} \sigma(\delta + \delta_D)
		\nonumber \\
	&=& \frac{2\gamma_0\Gamma}{\Omega_c^2}
		+\frac{2\Gamma^2}{\Omega_c^4}\Gamma_D^2
		+\frac{4\Gamma^2\delta^2}{\Omega_c^4},
\label{eq:sigmabar}
\end{eqnarray}
where $\Gamma_D$ is the $e^{-1}$ half width of the Gaussian distribution of $\delta_D$. For atoms characterized a temperature $T$ and a mass $m$, we have 
\begin{equation}
	\Gamma_D = \Delta k \sqrt{\frac{2 k_B T}{m}}.
\label{eq:GammaD}
\end{equation}
Here $k_B$ is the Boltzmann constant, and $\Delta k = |(\vec{k}_p - \vec{k}_c)\cdot \hat{z}|$ in the $\Lambda$-EIT system or $\Delta k = |(\vec{k}_p + \vec{k}_c)\cdot \hat{z}|$ in the Rydberg-EIT system, with $\vec{k}_p$ and $\vec{k}_c$ being the wave vectors of the probe and coupling fields. Note that $\Gamma_D$ can be negligible in the $\Lambda$-EIT system, because the two fields have very similar wavelengths and thus $\Delta k \approx 0$ in the co-propagation configuration of the probe and coupling fields. On the other hand, $\Gamma_D$ can be significant in the Rydberg-EIT system, because the two fields have rather different wavelengths.

The $\sigma (\delta)$ in Eq.~(\ref{eq:t}) is now replaced by $\bar{\sigma}(\delta)$ of Eq.~(\ref{eq:sigmabar}). Although the frequencies of the coupling and probe fields are locked to the resonance frequency of the two-photon transition, the frequency fluctuation randomly introduces a two-photon detuning, $\delta_f$, to the EIT system. We assume that the random fluctuation has a Gaussian distribution with the $e^{-1}$ half width of $\Gamma_f$. The average of transmission due to the Gaussian distribution is 
\begin{equation}
	\bar{t}(\delta) = \int_{-\infty}^{\infty}d\delta_f \;
		\frac{e^{-(\delta_f/\Gamma_f)^2}}{\sqrt{\pi}\Gamma_f} t(\delta+\delta_f).
\label{eq:tbar}
\end{equation}
Since reduction of the transmission is equivalent to an increment of the decoherence rate, one can define an effective decoherence rate $\gamma$ such that 
\begin{equation}
	\bar{t}(0) \equiv \exp(-2\alpha\gamma\Gamma/\Omega_c^2). 
\end{equation}
After evaluating Eq.~(\ref{eq:tbar}) at the EIT peak to get $\bar{t}(0)$, we obtain
\begin{equation}
	\gamma = \gamma_0 +\gamma_f + \gamma_D,
\label{eq:gamma}
\end{equation}
where
\begin{eqnarray}
\label{eq:gammaf}
	\gamma_f &\equiv& \frac{\Omega_c^2}{4\alpha\Gamma}
		\ln \left( 1+ \frac{4 \alpha \Gamma^2}{\Omega_c^4} \Gamma_f^2 \right), \\
\label{eq:gammaD}
	\gamma_D &\equiv& \frac{\Gamma}{\Omega_c^2} \Gamma_D^2.
\end{eqnarray}
Therefore, the total effective decoherence rate $\gamma$ consists of three parts: the intrinsic decoherence rate of the system $\gamma_0$, the frequency fluctuation-induced decoherence rate $\gamma_f$, and the Doppler shift-induced decoherence rate $\gamma_D$. Please note again that $\Gamma_f$ in Eq.~(\ref{eq:gammaf}) is the $e^{-1}$ half width of the Gaussian distribution of frequency fluctuation, and $\Gamma_D$ in Eq.~(\ref{eq:gammaD}) is that of Doppler shift.

\section{Experiment of $\Lambda$-type EIT}

We utilized the $\Lambda$-EIT system to verify the formula of the frequency fluctuation-induced decoherence rate $\gamma_f$ as shown in Eq.~(\ref{eq:gammaf}). The experiment was performed with the cigar-shaped cloud of cold $^{87}$Rb atoms produced by a magneto-optical trap (MOT) \cite{OurCigarMOT}. We optically pumped all population to a single Zeeman state of $|5S_{1/2},F=1,m_F=1\rangle$ before any measurement  \cite{OurPRL2013}. In the experiment, both of the probe and coupling and fields were $\sigma_+$-polarized. As shown in Fig.~1(a), only the Zeeman states $|1\rangle$,  $|2\rangle$, and $|3\rangle$ in the levels of $|5S_{1/2}\rangle$ and $|5P_{3/2}\rangle$ were relevant, which can avoid the complexity of multiple EIT subsystems \cite{OurOL2006, OurOE2009}. The wavelengths of probe and coupling fields were all around 780 nm, and their propagation directions were separated by a small angle of about 0.3$^{\circ}$. Thus, $\Gamma_D/(2\pi) \approx $ 1.8 kHz and the Doppler shift-induced decoherence rate $\gamma_D$ is negligible. 

\FigTwo

The experimental setup is shown in Fig.~2. A homemade external-cavity diode laser (ECDL) served as the master laser. We stabilized the ECDL's frequency by the scheme of saturated-absorption spectroscopy. The time constant of feedback loop in the frequency stabilization system was about 3 ms. Since the coupling and probe lasers were seeded or injection-locked by the light beams from the ECDL, their frequency difference was fixed with a high-degree stability. An electro-optic modulator generated 6.8 GHz sidebands in the ECDL beam, and the upper sideband locked the probe laser frequency. Acousto-optic modulators (AOMs) were used to switch the coupling field, generate Gaussian pulses of the probe field, and shift the frequencies of the two fields. As the probe and coupling fields interacted with the atoms, their $e^{-2}$ diameters were 0.30 and 4.4 mm, respectively. We set the maximum Rabi frequency of the probe field to about 0.036$\Gamma$, which is enough weak to be treated as the perturbation in the theoretical model. A photo-multiplier tube detected the probe light, and its output voltage was recorded by an oscilloscope (Agilent MSO6014A). All the experimental data presented in the paper were averaged for 512 times by the oscilloscope. 

\FigThree

The experimental parameters of coupling Rabi frequency ($\Omega_c$), optical depth or OD ($\alpha$), and decoherence rate ($\gamma$) were determined in the way illustrated by the example in Fig.~3. First, we measured the separation distance between two transmission minima, i.e. the Autler-Townes splitting, to determine $\Omega_c$ as shown in Fig.~3(a). According to Eq.~(\ref{eq:rho31}), the two minima occur at $\delta_{\pm} = (\Delta_c \pm \sqrt{\Delta_c^2+\Omega_c^2})/2$, where $\Delta_c$ is the one-photon detuning of the coupling field. At $\Delta_c \ll \Omega_c$, $\delta_+ - \delta_- \approx \Omega_c + \Delta_c^2/(2\Omega_c)$. Since we carefully minimized $\Delta_c$ in the measurement, the correction term $\Delta_c^2/(2\Omega_c)$ was about 1.3$\times$$10^{-4}$$\Gamma$ in Fig.~3(a). Here, the OD was intentionally reduced such that the two minima can be clearly observed. We swept the probe frequency by varying the rf frequency of AOM1 shown in Fig.~2. The sweeping rate was 240 kHz/$\mu$s, which is slow enough not to cause the transient effect \cite{OurJOSAb2004}. Asymmetry of the spectrum was caused by the decay of OD during the frequency sweeping. The asymmetry is not a problem, because the value of $\Omega_c$ determined in the low-to-high frequency sweeping differed from that in the high-to-low frequency sweeping merely by about 4\%. Knowing the vaue of $\Omega_c$, we then measured the delay time ($\tau_d$) to determine $\alpha$ as shown in Fig.~3(b), according to $\tau_d = \alpha\Gamma/\Omega_c^2$~\cite{Fleischhauer2005, OurPRA2006}. A short input pulse with the $e^{-1}$ full width of 3.5 $\mu$s was employed such that the delay time can be determined accurately. Knowing the values of $\Omega_c$ and $\alpha$, we finally measured the peak transmission of output pulse ($T_{\rm max}$) to determine $\gamma$ as shown in Fig.~3(c), according to $T_{\rm max} = \exp(-2\alpha\gamma\Gamma/\Omega_c^2)$. A long input pulse with the $e^{-1}$ full width of 35 $\mu$s at the two-photon resonance was employed. Once the values of $\Omega_c$, $\alpha$, and $\gamma$ were determined, we further calculated the predictions by numerically solving Eqs.~(1)-(3), and compared the short-pulse data with the predictions (similar to Fig.~2(a) in Ref.~\cite{OurPRL2013}). The good agreement between the experimental data and theoretical predictions demonstrates that the values of $\Omega_c$, $\alpha$, and $\gamma$ are convincing.

To verify Eq.~(\ref{eq:gammaf}), we controllably introduced fluctuation to the probe frequency via the AOM1 in Fig.~2. In the experiment, the probe pulse was the first-order beam of AOM1. The frequency of the first-order beam can be varied by the modulation voltage of the driver of AOM1. We employed a function generator to produce the voltage of Gaussian noise. The noise was added to the modulation voltage. Thus, the first-order beam, i.e., the probe pulse, possessed the Gaussian-distribution frequency fluctuation. Note that the amplitude noise (or power fluctuation) of the probe field, caused by the largest frequency fluctuation in this study, had the standard deviation less than 0.5\% of the mean power, which plays a negligible role in the decoherence rate. The center frequency of AOM1 made the two-photon transition resonant. Amplitude of the frequency fluctuation was determined by the beat note between the first-order and zeroth-order beams of AOM1. A photo detector (New Focus 1801) detected the beat note and its output signal was sent to a spectrum analyzer (Agilent EXA N9010A). Figure~4(a) shows representative the beat-note spectra measured by the spectrum analyzer. We fitted each spectrum with a Gaussian function. Since the beat note is proportional to the electric field of the first-order beam, $\Gamma_f$ is equal to the $e^{-1}$ half width of the best fit divided by $\sqrt{2}$.

\FigFour

The frequency fluctuation-induced decoherence rate $\gamma_f$ is the difference between the decoherence rates $\gamma$ with and without the frequency fluctuation. Value of $\gamma$ was determined by the method depicted in Fig.~3(c). In Fig.~4(b), the red squares are the experimental data of $\gamma_f$ as a function of $\Gamma_f$, which is the $e^{-1}$ half width of the Gaussian distribution of frequency fluctuation. In Fig.~4(c), the green, red, and blue squares represent the experimental data of $\gamma_f$ as functions of the OD. We subtracted $\gamma_0$ [$(4.6\pm1.4)$$\times$$10^{-4}$$\Gamma$] from $\gamma$ to obtain $\gamma_f$ in the above measurements. The black lines in Figs.~4(b) and 4(c) represent the theoretical predictions according to Eq.~(\ref{eq:gammaf}), where the calculation parameters of coupling Rabi frequency, OD, and $\Gamma_f$ were determined by the methods illustrated in Figs.~3(a), 3(b), and 4(a). The good agreement between the experimental data and the theoretical predictions demonstrates Eq.~(\ref{eq:gammaf}) is valid. 

\section{Experiment of Rydberg-State EIT}

We now study whether Eq.~(\ref{eq:gamma}) can quantitatively describe the decoherence rate in the Rydberg-EIT system. The experimental study was carried out with the cold $^{87}$Rb atoms trapped by the same MOT \cite{OurCigarMOT}. We optically pumped all population to a single Zeeman state of $|5S_{1/2},F=2,m_F=2\rangle$ before any measurement \cite{OurNComms2014}. In the experiment, both of the probe and coupling and fields were $\sigma_+$-polarized. As shown in Fig.~1(b), only the Zeeman states $|1\rangle$,  $|2\rangle$, and $|3\rangle$ in the levels of $|5S_{1/2}\rangle$, $|5P_{3/2}\rangle$, and $|32D_{5/2}\rangle$ were relevant, which can avoid the complexity of multiple EIT subsystems \cite{OurOL2006, OurOE2009}. We selected a low principle quantum number of $n = 32$ for the Rydberg state such that the DDI effect can be negligible in this work according to our estimation \cite{Saffman2008, OurEstimation}. The wavelengths of probe and coupling fields were around 780 nm and 482 nm, respectively. They propagated in the opposite directions to minimize the Doppler effect \cite{Xiao1995}. As the probe and coupling fields interacted with the atoms, their $e^{-2}$ diameters were 300 and 350 $\mu$m, respectively. We set the maximum Rabi frequency of the probe field to about 0.034$\Gamma$.

\FigFive

The setup of Rydberg-EIT experiment is depicted in Fig.~5. An ECDL (Toptica DLC DL pro) injection-locked or seeded the probe laser. We stabilized the ECDL's frequency by using the Pound-Drever-Hall (PDH) scheme in the saturated-absorption spectrum measured with a hot atomic vapor cell. The bandwidth of feedback loop for the probe laser in the frequency stabilization system was about 4 MHz. The coupling field was generated by the laser system of Toptica TA-SHG pro. We stabilized the frequency of the coupling laser by using the PDH scheme in the EIT spectrum, in which light beams from the ECDL and the coupling laser interacted with the atomic vapor in another hot vapor cell \cite{Adams2009}. The bandwidth of feedback loop for the coupling laser in the frequency stabilization system was about 50 kHz. We used the AOM3 in Fig.~5 to make the probe frequency seen by the cold atoms resonant to or detuned away from the transition frequency of $|1\rangle$ $\rightarrow$ $|3\rangle$. Since the coupling frequency was kept resonant to the transition frequency of $|3\rangle$ $\rightarrow$ $|2\rangle$, the AOM3 also set the two-photon detuning in the measurement. When the probe and coupling frequencies were both locked, we used an independent PDH signal (not used in the frequency locking) at the bandwidth of 1 MHz to determine the frequency fluctuations of the probe and coupling lasers. The root-mean-square value of the total frequency fluctuation is 150 kHz, indicating $\Gamma_f/(2\pi) =$ 210 kHz. 

\FigSix

We measured the atom temperature with the Rydberg-EIT light storage to determine $\Gamma_D$, which is the $e^{-1}$ half width of the Gaussian distribution of Doppler shift. The main plot in Fig.~6 shows retrieval efficiency (ratio of output to input energies) of the probe pulse as a function of the storage time. The best fit of the data indicates that the coherence time or $e^{-1}$ decay time, $\tau_{\rm coh}$, is 1.1 $\mu$s. Since $\tau_{\rm coh}^{-1} =  |\vec{k}_p + \vec{k}_c| \sqrt{k_B T/m}$ \cite{Pan2009, OurJPB2011} and $|\vec{k}_p + \vec{k}_c|$ = 5.0$\times$$10^6$ m$^{-1}$ in our case, the atom temperature was about 350~$\mu$K in the Rydberg-EIT experiment. In another measurement of the $\Lambda$-EIT light storage, $\tau_{\rm coh}$ was 125 $\mu$s or the atom temperature was around 350~$\mu$K. The two values of atom temperature determined by the Rydberg-EIT and $\Lambda$-EIT light storages are consistent. According to the atom temperature of 350~$\mu$K and Eq.~(\ref{eq:GammaD}), we can know $\Gamma_D/(2\pi) =$ 200~kHz.

\FigSeven

The experimental parameters in the Rydberg-EIT system were determined in the similar way as those in the $\Lambda$-EIT system. Figure~7 shows an example of the determination procedure. In Fig.~7(a), the sweep of probe frequency was done by AOM3 in Fig.~5. In Fig.~7(b), a short input pulse with the $e^{-1}$ full width of 0.53 $\mu$s was employed. In Fig.~7(c), we used a frequency counter (Agilent 53131A), monitoring the rf frequency of the AOM3, to determine the two-photon detuning $\delta$.

To determine $\gamma$, we fitted the data points in Fig.~7(c) with the fitting function given by
\begin{equation}
	\exp \left[ -\frac{2\alpha \gamma \Gamma}{\Omega_c^2} 
		 -\frac{4\alpha\Gamma^2 \delta^2}{\Omega_c^4}
		-\frac{16\alpha\Gamma^2 (2\Omega_c^2-\Gamma^2) \delta^4}{\Omega_c^8} \right].
\label{eq:tvdetuning}
\end{equation}
Since the absorption cross section in Eq.~(\ref{eq:sigma}) is valid only around the peak of the EIT spectrum, we need to add the term of $\delta^4$ in the fitting function. The above function is derived by expanding $\rho_{31}$ of Eq.~(\ref{eq:rho31}) up to $\delta^4$ to obtain the absorption cross section $\sigma(\delta)$, and performing the integrals of Eq.~(\ref{eq:sigmabar}) and then Eq.~(\ref{eq:tbar}). The terms of $(\Gamma_f/\Omega_c)^n$ and $(\Gamma_D/\Omega_c)^n$ with $n \geq 4$ are dropped during the derivation. To fit the data points in Fig.~7(c), $\Omega_c$ and $\alpha$ are fixed to the values determined in Figs.~7(a) and 7(b). The maximum transmission of the best fit determines the value of $\gamma$. We further calculated the predictions by numerically solving Eqs.~(1)-(3) with the values of $\Omega_c$, $\alpha$, and $\gamma$ determined in the above, and compared the short-pulse data with the predictions. The good agreement between the data and predictions makes the values of coupling Rabi frequency $\Omega_c$, optical depth $\alpha$, and decoherence rate $\gamma$ more convincing.

\FigEight

We next studied whether Eq.~(\ref{eq:gamma}) can quantitatively describe the decoherence processes in our Rydberg-EIT system. The decoherence rates $\gamma$ at various coupling Rabi frequencies $\Omega_c$ were measured. Figure 8 shows $\gamma$ as a function of $1/\Omega_c^2$. The red squares are the experimental data determined by the method illustrated in Fig.~7(c). The black lines in Fig.~8 are the theoretical predictions calculated from Eq.~(\ref{eq:gamma}). In the calculation, we used the values of $\Omega_c$ and optical depth ($\alpha$) determined in Figs.~7(a) and 7(b), and set $\Gamma_f/(2\pi) = 210$~kHz, $\Gamma_D/(2\pi) = 200$~kHz, and $\gamma_0/(2\pi) = 22$~kHz. The value of $\Gamma_f$ was determined by the demodulated PDH signals in the probe and coupling frequency stabilization systems. The value of $\Gamma_D$ was determined by the light storage measurement. The value of $\gamma_0$ minimizes the difference between the data points and the theoretical predictions. Please note that under $\Omega_c^2/(\Gamma\sqrt{\alpha}) \gg 2\Gamma_f$ where $\Omega_c^2/(\Gamma\sqrt{\alpha})$ is the $e^{-1}$ full width of EIT window, Eq.~(\ref{eq:gammaf}) and Eq.~(\ref{eq:gamma}) become
\begin{eqnarray}
	\gamma_f &\approx&  \frac{\Gamma}{\Omega_c^2}\Gamma_f^2, \\
	\gamma &\approx& \gamma_0 + \frac{\Gamma}{\Omega_c^2}(\Gamma_D^2+\Gamma_f^2).
\end{eqnarray}
Hence, $\gamma$ is linearly proportional to $1/\Omega_c^2$ and becomes independent of $\alpha$ at large values of $\Omega_c$ as shown in Fig.~8. The good agreement between the data and predictions shows that the theoretical model or Eq.~(\ref{eq:gamma}) in this work describes the decoherence processes in the Rydberg-EIT system very well. 

After the amount of frequency fluctuation due to the laser frequency stabilization system was confirmed in Fig.~8, we further studied the decoherence rate, $\gamma$, as a function of the frequency fluctuation, $\Gamma_f$, as shown in Fig.~9. The additional frequency fluctuation was introduced via the AOM3 in Fig.~5 in the same way as we described in Sec.~III. The amplitude noise (or power fluctuation) of the probe field, caused by the largest frequency fluctuation introduced to the AOM3, had the standard deviation of about 0.4\% of the mean power, which plays a negligible role in the decoherence rate. Now, the value of $\Gamma_f$ is equal to $\sqrt{\Gamma_{f,{\rm laser}}^2+\Gamma_{f,{\rm AOM}}^2}$, where $\Gamma_{f,{\rm laser}}$ is the frequency fluctuation due to the stabilization system, and $\Gamma_{f,{\rm AOM}}$ is the frequency fluctuation due to the AOM3. We employed a Gaussian input pulse with the $e^{-1}$ full width of 7.0~$\mu$s in the measurement, and determined the value of $\gamma$ by the transmission of the output pulse. In Fig.~9, the consistency between the experimental data and theoretical prediction is satisfactory.

\FigNine

Finally, one might worry that the separation distance between two transmission minima, $\Delta\omega_{\rm AT}$, in Fig.~7(a) and the delay time, $\tau_d$, in Fig.~7(b) can be influenced by the effects of frequency fluctuation and atomic motion. The two effects, i.e., $\Gamma_f$ and $\Gamma_D$, were switched off or negligible in Figs.~3(a) and 3(b), but must be present in Figs.~7(a) and 7(b). Using  Eqs.~(\ref{eq:rho31}), (\ref{eq:sigmabar}), and (\ref{eq:tbar}), we derive the values of $\Delta\omega_{\rm AT}$ and $\tau_d$ in the presence of $\Gamma_f$ and $\Gamma_D$. It can be shown that
\begin{eqnarray}
	\Delta\omega_{\rm AT} &=& \Omega_c 
		\left[ 1 +\frac{3(\Gamma_f^2+\Gamma_D^2)}{\Omega_c^2} \right], \\
	\tau_d &=& \frac{\alpha \Gamma}{\Omega_c^2} \left[1 
		+\frac{6(\Gamma_f^2+\Gamma_D^2)(\Omega_c^2-\Gamma^2)}{\Omega_c^4} \right].
\end{eqnarray}
In our system, $\Gamma_f = 2\pi \times 210$~kHz and $\Gamma_D = 2\pi\times 200$~kHz. The minimum value of $\Omega_c$ in this study was $0.71$$\Gamma$ or $2\pi\times 4.3$~MHz. Therefore, $\Delta\omega_{\rm AT} = \Omega_c$ and $\tau_d = \alpha \Gamma/\Omega_c^2$ used in Figs.~7(a) and 7(b) are the good approximations. 


\section{Conclusion}

In this work, we systematically studied the effect of laser frequency fluctuation on the decoherence rate of the $\Lambda$-type and Rydberg-state EIT systems. The laser frequency fluctuation randomly introduces a two-photon detuning to the system, resulting in attenuation of the probe field. The attenuation is equivalent to an increment of the decoherence rate. Using the steady-state solution of the optical Bloch equation and the Maxwell-Schr\"{o}dinger equation, we derived the formula in Eq.~(\ref{eq:gammaf}) to describe the frequency fluctuation-induced decoherence rate, $\gamma_f$. The formula of $\gamma_f$ was tested in the $\Lambda$-EIT system with the laser-cooled $^{87}$Rb atoms. The experimental data of $\gamma_f$ are in the good agreement with the theoretical predictions of Eq.~(\ref{eq:gammaf}). 

We further studied the decoherence processes in the Rydberg-EIT system, in which moving atoms induce non-negligible two-photon detunings due to the Doppler shift. We considered the distribution of Doppler shift among the atoms, and obtained the formula in Eq.~(\ref{eq:gammaD}) to describe the Doppler shift-induced decoherence rate, $\gamma_D$. The total effective decoherence rate $\gamma$ shown in Eq.~(\ref{eq:gamma}) consists of three parts: $\gamma_f$, $\gamma_D$, and $\gamma_0$, the latter $\gamma_0$ being an intrinsic decoherence rate in the system. The experimental study of $\gamma$ was carried out with the cold atoms of 350 $\mu$K. We utilized the Rydberg state of $|32D_{5/2}\rangle$, in which the dipole-dipole interaction can be neglected. A rather low value of $\gamma$ of $2\pi\times$48~kHz was achieved at a moderate coupling Rabi frequency ($\Omega_c$) of $2\pi\times$4.3~MHz. The experimental data of $\gamma$ are consistent with the theoretical predictions. 

According to Eqs.~ (\ref{eq:gamma}), (\ref{eq:gammaf}), and (\ref{eq:gammaD}), larger values of $\Omega_c$ make smaller $\gamma$. Furthermore, as the EIT linewidth is much greater than the amplitude of the frequency fluctuation, $\gamma$ is linearly proportional to $1/\Omega_c^2$ and asymptotically approaches to $\gamma_0$. In our Rydberg-EIT system, $\gamma_0$ is approximately $2\pi\times$22 kHz, which comes from the natural linewidth of the Rydberg state, the Lorentzian-type laser linewidths of the probe and coupling fields, and other decoherence processes. The results of our work are useful for the estimation of outcomes or decoherence rates of Rydberg-EIT experiments, and provide a better understanding of the Rydberg-EIT effect.

\section*{Acknowledgments}
This work was supported by the Ministry of Science and Technology of Taiwan under Grant Nos.~105-2923-M-007-002-MY3, 106-2119-M-007-003, and 107-2745-M-007-001, and the project TAP LLT-2/2016 of the Research Council of Lithuania. JR and GJ also acknowledge a support by the National Center for Theoretical Sciences, Taiwan. All authors thank Dr.~Ming-Shien Chang for lending us the laser system of Toptica TA-SHG pro.


\end{document}